\documentclass[a4paper]{article}
\usepackage{amsmath,amssymb}
\begin{document}
\begin{center}
\Large{ Bilinear Equations and B\"acklund Transformation for Generalized Ultradiscrete Soliton Solution}\\
\vspace{1cm}
\large{Hidetomo Nagai\footnote{e-mail hdnagai@aoni.waseda.jp}, Daisuke Takahashi\footnote{e-mail daisuket@waseda.jp}}\\
Faculty of Science and Engineering, Waseda University, 3-4-1, Okubo, Shinjuku-ku, Tokyo 169-8555, Japan
\end{center}
\begin{abstract}
  Ultradiscrete soliton equations and B\"acklund transformation for a generalized soliton solution are presented.  The equations include the ultradiscrete KdV equation or the ultradiscrete Toda equation in a special case.  We also express the solution by the ultradiscrete permanent, which is defined by ultradiscretizing the signature-free determinant, that is, the permanent.  Moreover, we discuss a relation between B\"acklund transformations for discrete and ultradiscrete KdV equations.
\end{abstract}
%\maketitle
%%%%%%%%%%%%%%%%%%%%%%%%%%%%%%%%%%%%%%%%%%%%%%%%%%%%%%%%%%%%%%%%%%%%%
\section{Introduction}
%%%%%%%%%%%%%%%%%%%%%%%%%%%%%%%%%%%%%%%%%%%%%%%%%%%%%%%%%%%%%%%%%%%%%
Soliton equation has explicit $N$-soliton solutions and an infinite number of conserved quantities generally.  In the beginning of the development of the soliton theory, continuous or semi-discrete soliton equations were studied mainly.  For example, the Korteweg-de Vries (KdV) equation is a continuous soliton equation of PDE type, and the Toda equation is a semi-discrete soliton equation with continuous and discrete independent variables.  There are two types of soliton solutions to the bilinear equations derived from these equations.  One is expressed by a sum of a finite number of exponential functions, which was first proposed by Hirota\cite{Hirotacont1, Hirotacont2}.  We call this type of expression Type I.  The other is expressed by Wronski determinant \cite{Satsuma, Freeman}.  We call this type of expression Type II.\par
  After the discovery of various continuous or semi-discrete soliton equations, discrete soliton equations of which independent variables are all discrete were proposed\cite{Hirotadis1, Hirotadis4}.  Discrete soliton equation is also transformed into the bilinear equation and has multi-soliton solutions. It has also two types of expressions, Type I and II, where the determinant of Type II is generally the Casorati determinant for discrete soliton equations.\par
  Discretization process is completed if dependent and independent variables are all discretized.  In the 1990s, Tokihiro {\it et al.} proposed the ultradiscretization method to discretize dependent variables\cite{Tokihiro}.  The key formula in the method is
\begin{equation}
  \lim _{\varepsilon \to +0}\varepsilon \log (e^{a/\varepsilon }+e^{b/\varepsilon }) = \max (a, b).
\end{equation}
Usual addition, multiplication and division for the real values in the original discrete equation are replaced with max operation, addition and subtraction respectively by this method.  Due to these replacements, dependent variables can be discrete in the ultradiscrete equation if we use appropriate constants and initial values.  Many ultradiscrete soliton equations or cellular automata have been proposed and the integrability is shown even for the digitized equations\cite{Matsukidaira, Tsujimoto}.\par
  However, the operation in the ultradiscrete equation corresponding to the subtraction in the discrete equation is not well-defined.  Thus we can not ultradiscretize a discrete equation automatically.  This obstruction is called a `negative problem'\cite{uKdV, Nagai}.  Thus the above soliton solution of Type II can not be ultradiscretized directly since the antisymmetry is crucial for the determinant.  On the other hand, the solution of Type I can be ultradiscretized generally choosing the appropriate parameters included in the solution.\par
  The imbalance between the two types of expression for the ultradiscrete soliton solution is partially solved.  One of the authors (Takahashi) and Hirota proposed the ultradiscrete analogue of determinant solution for the ultradiscrete KdV (uKdV) equation\cite{uKdV}.  One of the authors (Nagai) proposed the similar type of solution for the ultradiscrete Toda (uToda) equation\cite{Nagai}.  This analogue is called an `ultradiscrete permanent' (UP) defined by
\begin{equation} 
\max[a_{ij}]\equiv \max_{\pi} \sum_{1\le i\le N}a_{i\pi _i},
\end{equation}
where $[a_{ij}]$ denotes an arbitrary $N\times N$ matrix and $\pi=\{\pi_1, \pi_2, \dots, \pi_N\}$ an arbitrary permutation of 1, 2, $\dots$, $N$.  The $N$-soliton solution to the ultradiscrete bilinear equation of uKdV equation is expressed by the following two forms, 
\begin{equation} \label{solution kdv I}
  f^n_i = \max_{\mu_j=0, 1}\Bigl(\sum_{1\le j\le N} \mu_j s_j(n, i) -\sum_{1\le j<j'\le N} \mu_j\mu_{j'} a_{jj'}\Bigr),
\end{equation}
and
\begin{equation}  \label{solution kdv II}
  \tilde{f}^n_i =\frac{1}{2}\max 
\begin{bmatrix}
  |s_1(n-N+1, i)| & |s_1(n-N+3, i)| & \dots &|s_1(n+N-1, i)|\\
  \dots & \dots &  \dots &\dots\\
  |s_N(n-N+1, i)| & |s_N(n-N+3, i)| & \dots &|s_N(n+N-1, i)|
\end{bmatrix},
\end{equation}
where $\max_{\mu_j=0,1}X(\mu_1,\mu_2,\ldots,\mu_N)$ denotes the maximum value of $X$ in $2^N$ possible cases of $\{\mu_1,\mu_2,\ldots,\mu_N\}$ replacing each $\mu_j$ by $0$ or $1$.  We call the form (\ref{solution kdv I}) Type I and (\ref{solution kdv II}) Type II respectively in this article.\par
B\"acklund transformation is an important object in the soliton theory since it gives the links among equations or solutions\cite{Hirotadis1, Hirotadis4}.  The ultradiscrete version of the B\"acklund transformation is discussed in \cite{Shinzawa} or \cite{uBack}.  The equations treated in the references are the ultradiscrete Kadomtsev-Petviashvili equation and the uKdV equation. \par
In this article, we consider a generalized solution of both types,
\begin{equation}  \label{solution I}
  f_i^n=\max_{\mu_j=0,1}\Bigl(\sum_{1\le j\le N}\mu_js_j(n,i)-\sum_{1\le j<j'\le N}\mu_j\mu_{j'}r_j\Bigr)
\end{equation}
and 
\begin{equation} \label{solution II}
\tilde f^n_i = \max \begin{bmatrix}
   |s_1+(-N+1)\displaystyle\frac{r_1}{2}| & |s_1+(-N+3)\displaystyle\frac{r_1}{2}| &\dots & |s_1+(N-1)\displaystyle\frac{r_1}{2}| \\ 
   \dots & \dots &\dots & \dots \\
   |s_N+(-N+1)\displaystyle\frac{r_N}{2}| & |s_N+(-N+3)\displaystyle\frac{r_N}{2}| &\dots & |s_N+(N-1)\displaystyle\frac{r_N}{2}| 
\end{bmatrix},
\end{equation}
where
\begin{equation}  \label{def spr}
\begin{aligned}  
  & s_j=s_j(n,i)=p_jn-q_ji+c_j, \qquad 0\le p_1\le p_2\le \dots\le p_N,\\
  & r_j=kp_j+lq_j, \qquad \text{$k$, $l$: non-negative constants}.
\end{aligned}
\end{equation}
We discuss ultradiscrete soliton equations and a B\"acklund transformation for this solution.\par
  The contents of this article are as follows. In Section 2, we give equations which $f^n_i$ and $\tilde f^n_i$ satisfy under some conditions.  In Section 3, we show a B\"acklund transformation between $N$ and $(N+1)$-soliton solutions.  In Section 4, we ultradiscretize the B\"acklund transformations for the discrete KdV equation and obtain those for the uKdV equation.  The results suggests an algebraic correspondence between the solutions of determinant and of UP, In Section 5, we give concluding remarks.\par
  The proofs in this article are given under the condition $p_j\ge 0$ for simplicity.  However, note that these results can be easily extended to the case of arbitrary $p_j$'s by replacing (\ref{solution I}) with 
\begin{equation}  
  f_i^n=\max_{\mu_j=0,1}\Bigl(\sum_{1\le j\le N}\mu_js_j(n,i)-\sum_{1\le j<j'\le N}\mu_j\mu_{j'}a_{jj'}\Bigr),
\end{equation}
where
\begin{equation}
  a_{jj'} = \min (\max (r_j, -r_{j'}), \max (-r_j, r_{j'})).
\end{equation}
%%%%%%%%%%%%%%%%%%%%%%%%%%%%%%%%%%%%%%%%%%%%%%%%%%%%%%%%%%%%%%%%%%%%%
\section{Bilinear Equations for the Generalized Soliton Solution}
%%%%%%%%%%%%%%%%%%%%%%%%%%%%%%%%%%%%%%%%%%%%%%%%%%%%%%%%%%%%%%%%%%%%%
  First we give the following propositions.
\newtheorem{proposition}{Proposition}
\begin{proposition}  \label{prop 1}
  The generalized solution (\ref{solution I}) with a dispersion relation
\begin{equation}  \label{disp kdv}
  q_j=\min(p_j,L),
\end{equation}
where $L$ ($\ge0$) is a constant, satisfies a bilinear equation
\begin{equation}  \label{kdv}
  f^n_i+ f^{n+k}_{i-l+1} = \max ( f^{n+k-1}_{i-l}+ f^{n+1}_{i+1},  f^{n+k}_{i-l}+ f^n_{i+1}-L) \qquad (k\ge 2).
\end{equation}
\end{proposition}
Similarly, the following proposition holds.
\begin{proposition}  \label{prop 2}
  The generalized solution (\ref{solution I}) with a dispersion relation
\begin{equation}  \label{disp toda}
  q_j=\max(0,p_j-L),
\end{equation}
satisfies
\begin{equation}  \label{toda}
  f^n_i+ f^{n+k-1}_{i-l-1} = \max ( f^{n+k-1}_{i-l}+ f^n_{i-1},  f^{n+k}_{i-l}+ f^{n-1}_{i-1}-L) \qquad (k, l \ge 1).
\end{equation}
\end{proposition}
Note that (\ref{kdv}) reduces to the uKdV equation in the case of $(k,l)=(2,0)$, and (\ref{toda}) reduces to the uToda equation in the case of $(k, l)=(1,1)$ respectively.\par
  Since the whole of proof of the propositions is long, let us show the outline.  First, considering $f^{n+x}_{i+y} + f^{n+z}_{i+w}$ with arbitrary constants $x$, $y$, $z$ and $w$,  we have from (\ref{solution I})
\begin{equation} \label{equation 2-1}
\begin{aligned}
  &f^{n+x}_{i+y} + f^{n+z}_{i+w} = \max_{\mu_j, \nu_j= 0, 1}\Bigl( \sum_{1\le j\le N} (\mu_j +\nu_j) s_j\\
  & +\sum_{1\le j\le N}(\mu_j(x p_j- y q_j)+\nu_j(zp_j- w q_j)) -  \sum_{1\le j<j'\le N} (\mu_j \mu_{j'}  +\nu_j \nu_{j'}) r_j  \Bigr)
\end{aligned}
\end{equation}
where $s_j$ denotes $s_j(n, i)$ for short.  Using new parameters $\lambda_j$ and $\sigma_j$ defined by $\lambda_{j}=\mu_{j}+\nu_{j}$, $\sigma_j=\mu_j-\nu_j$ ($1\le j\le N$), (\ref{equation 2-1}) reduces to
\begin{equation}  \label{equation 2-2}
\begin{aligned}
  &f^{n+x}_{i+y} + f^{n+z}_{i+w}\\
=&\max_{(\lambda_j, \sigma_j)}\Bigl( \sum_{1\le j\le N} \lambda_j s_j  +\frac{1}{2}\sum_{1\le j\le N}\lambda_j ((x+z) p_j-(y+w) q_j)\\
  &+\frac{1}{2}\sum_{1\le j\le N}\sigma_j ((x-z)p_j -(y-w )q_j) -\frac{1}{2}\sum_{1\le j<j'\le N} (\lambda_j \lambda_{j'}  +\sigma_j \sigma_{j'}) r_j \Bigr).
\end{aligned}
\end{equation}
Note that the pair $(\lambda_j,\sigma_j)$ can be one of the following,
\begin{equation}
  (0, 0),\quad (1,1),\quad (1,-1),\quad (2, 0),
\end{equation}
and $\max_{(\lambda_j, \sigma_j)}X(\lambda_1,\ldots,\lambda_N,\sigma_1,\ldots,\sigma_N)$ denotes the maximum value of $X$ in $4^N$ possible cases of $\{\lambda_1,\ldots,\lambda_N,\sigma_1,\ldots,\sigma_N\}$ replacing each $(\lambda_j,\sigma_j)$ by one of the above four pairs.  
Comparing the terms in (\ref{kdv}) substituting (\ref{equation 2-2}), we can show that Proposition~\ref{prop 1} holds if the following is proved for any $N'$ satisfying $1\le N'\le N$ \cite{uKdV}.
\begin{equation}  \label{from prop 1}
\begin{aligned}
  \max_{\sigma_j=\pm 1}\Bigl(& \sum_{1\le j\le N'}\sigma_j (k p_j+(l-1) q_j)- \sum_{1\le j<j'\le N'}\sigma_j \sigma_{j'} r_j \Bigr) \\
  = \max \Bigl(& \max_{\sigma_j=\pm 1}\Bigl( \sum_{1\le j\le N'}\sigma_j((k-2)p_j+(l+1) q_j)- \sum_{1\le j<j'\le N'}\sigma_j \sigma_{j'} r_j \Bigr),\\
 & \max_{\sigma_j=\pm 1}\Bigl( \sum_{1\le j\le N'}\sigma_j (kp_j+(l+1) q_j)- \sum_{1\le j<j'\le N'}\sigma_j \sigma_{j'} r_j  \Bigr) -2L\Bigr).
\end{aligned}
\end{equation}
Using the dispersion relation (\ref{disp kdv}), the maxima of LHS and of the first argument in RHS of (\ref{from prop 1}) are both given by the case of $\sigma_j=(-1)^{N'-j}$ ($1\le j\le N'$)\cite{Nagai}.  About the second argument in RHS, the case of $\sigma_ {N'}=1$ and $\sigma_j = (-1)^{N'-j+1}$ $(1\le j\le N'-1)$ gives the maximum.  Therefore, (\ref{from prop 1}) becomes
\begin{equation}  \label{equation 2-2-2}
  \max(-p_{N'}+q_{N'}+p_{N'-1}-q_{N'-1}-p_{N'-2}+q_{N'-2}+\dots,\ q_{N'}-L)=0,
\end{equation}
after the above evaluation.  In particular, the dispersion relation (\ref{disp kdv}) derives
\begin{equation}
\begin{cases}
    -p_{N'}+q_{N'}+p_{N'-1}-q_{N'-1}-\dots \le 0\quad\text{and}\quad q_{N'}-L =0 &  \text{if $p_N\ge L$}, \\
    -p_{N'}+q_{N'}+p_{N'-1}-q_{N'-1}-\dots =0\quad\text{and}\quad q_{N'}-L <0 &  \text{if $p_N<L$}
\end{cases}.
\end{equation}
Thus (\ref{equation 2-2-2}) holds for any $1\le N'\le N$ and Proposition~\ref{prop 1} is proved.\par
  Comparing the terms in (\ref{toda}), we can show that Proposition~\ref{prop 2} holds if the following is proved for any $1\le N'\le N$.
\begin{equation}
\begin{aligned}
  \max_{\sigma_j=\pm 1}\Bigl(& \sum_{1\le j\le N'}\sigma_j ((k-1) p_j+(l+1) q_j)- \sum_{1\le j<j'\le N'}\sigma_j \sigma_{j'} r_j  \Bigr) \\
  = \max \Bigl(& \max_{\sigma_j=\pm 1}\Bigl( \sum_{1\le j\le N'}\sigma_j((k-1)p_j+(l-1) q_j)- \sum_{1\le j<j'\le N'}\sigma_j \sigma_{j'} r_j \Bigr) ,\\
 & \max_{\sigma_j=\pm 1}\Bigl( \sum_{1\le j\le N'}\sigma_j ((k+1)p_j+(l-1) q_j)- \sum_{1\le j<j'\le N'}\sigma_j \sigma_{j'} r_j  \Bigr) -2L\Bigr).
\end{aligned}
\end{equation}
Using the dispersion relation (\ref{disp toda}), the case of $\sigma_j=(-1)^{N'-j}$ ($1\le j\le N'$) gives the maxima of LHS and of the first argument in RHS.  The case of $\sigma_ {N'}=1$ and $\sigma_j = (-1)^{N'-j+1}$ $(1\le j\le N'-1)$ gives the maximum of the second argument in RHS.  Therefore, the above equation becomes
\begin{equation}
  \max(-q_{N'}+q_{N'-1}-q_{N'-2}+q_{N'-3}-\dots,\ p_{N'}-q_{N'}-L)=0,
\end{equation}
after the above evaluation.  This equation holds since 
\begin{equation}
\begin{cases}
  -q_{N'}+q_{N'-1}-q_{N'-2}+q_{N'-3}-\dots\le 0\quad\text{and}\quad p_{N'}-q_{N'}-L =0 &  \text{if $p_N\ge L$,}\\
  -q_{N'}+q_{N'-1}-q_{N'-2}+q_{N'-3}-\dots=0\quad\text{and}\quad p_{N'}-q_{N'}-L <0 &  \text{if $p_N<L$}
\end{cases}.
\end{equation}
Thus Proposition~\ref{prop 2} is proved.  \par
Moreover these propositions lead the following propositions.  
\begin{proposition}\label{prop 1-2}
  The generalized solution (\ref{solution II}) satisfies
\begin{equation}  \label{ukdv2}
  \tilde f_i^n+\tilde f_{i-l+1}^{n+k}=\max(\tilde f_{i-l}^{n+k-1}+\tilde f_{i+1}^{n+1},\,\tilde f_{i-l}^{n+k}+\tilde f_{i+1}^n-2L)
\end{equation}
under the dispersion relation (\ref{disp kdv}).
\end{proposition}
\begin{proposition}  \label{prop 2-2}
  The generalized solution (\ref{solution II}) satisfies
\begin{equation}  \label{toda2}
  \tilde f_i^n+\tilde f_{i-l-1}^{n+k-1}=\max(\tilde f_{i-l}^{n+k-1}+\tilde f_{i-1}^n,\,\tilde f_{i-l}^{n+k}+\tilde f_{i-1}^{n-1}-2L)
\end{equation}
under the dispersion relation (\ref{disp toda}).
\end{proposition}
Both propositions are proved by reducing (\ref{ukdv2}) and (\ref{toda2}) to (\ref{kdv}) and (\ref{toda}) respectively.  Using a formula\cite{uKdV}
\begin{equation}
\begin{aligned}  \label{formula}
  &\max \begin{bmatrix}
   |y_1+(-N+1)r_1| & |y_1+(-N+3)r_1| &\dots & |y_1+(N-1)r_1| \\ 
   \dots & \dots &\dots & \dots \\
   |y_N+(-N+1)r_N| & |y_N+(-N+3)r_N| &\dots & |y_N+(N-1)r_N| \\ 
  \end{bmatrix} \\
 =&\max_{\rho_j= \pm 1}\Bigl( \sum_{1\le j\le N}\rho_jy_j- \sum_{1\le j<j'\le N} \rho_j\rho_{j'} r_j \Bigr)+\sum_{1\le j<j'\le N} r_{j'},
\end{aligned}
\end{equation}
where
\begin{equation*}
  0\le r_1\le r_2\le \dots \le r_N, 
\end{equation*}
and the transformation $\rho_j=2\mu_j-1$, we have
\begin{equation*}
\begin{aligned}
   \tilde f^n_i=& \max_{\rho_j= \pm 1}\Bigl( \sum_{1\le j\le N}\rho_js_j- \frac{1}{2}\sum_{1\le j<j'\le N} \rho_j\rho_{j'} r_j \Bigr)+\frac{1}{2}\sum_{1\le j<j'\le N} r_{j'}  \\
\Bumpeq &\max_{\mu_j=0, 1}\Bigl( 2\sum_{1\le j\le N}\mu_js_j- 2\sum_{1\le j<j'\le N} \mu_j\mu_{j'} r_j +\sum_{1\le j<j'\le N}(\mu_j+\mu_{j'})r_j\Bigr)\\
  &\qquad -\sum_{1\le j\le N}s_j-\frac{1}{2}\sum_{1\le j<j'\le N}r_j\\
  \Bumpeq &2\max_{\mu_j=0, 1}\Bigl( \sum_{1\le j\le N}\mu_j(s_j+\frac{N-j}{2}r_j+\frac{1}{2}\sum_{1\le j'\le j-1}r_{j'}) - \sum_{1\le j<j'\le N} \mu_j\mu_{j'} r_j \Bigr)\\
   &\qquad-\sum_{1\le j\le N}s_j.
\end{aligned}
\end{equation*}
Here $f^n_i\Bumpeq g^n_i$ denotes that $f^n_i$ and $g^n_i$ give the same solution of (\ref{ukdv2}) or (\ref{toda2}).  Hence, using a replacement
\begin{equation}
  c_j + \frac{N-j}{2}r_j +\frac{1}{2}\sum_{1\le j'\le j-1} r_{j'}\to c_j,
\end{equation}
we have
\begin{equation}
   \tilde f^{n+x }_{i+y } \Bumpeq 2f_{i+y}^{n+x} -\sum_{1\le j\le N}s_j(n+x, i+y).
\end{equation}
Thus, (\ref{ukdv2}) is reduced to (\ref{kdv}) by adding $\sum_{1\le j\le N}(2s_j+kp_j+(l-1)q_j)$ to both sides, and (\ref{toda2}) is reduced to (\ref{toda}) by adding $\sum_{1\le j\le N}(2s_j+(k-1)p_j+(l+1)q_j)$ respectively.\par
%%%%%%%%%%%%%%%%%%%%%%%%%%%%%%%%%%%%%%%%%%%%%%%%%%%%%%%%%%%%%%%%%%%%%
\section{B\"acklund Transformation for the Generalized Soliton Solution}
%%%%%%%%%%%%%%%%%%%%%%%%%%%%%%%%%%%%%%%%%%%%%%%%%%%%%%%%%%%%%%%%%%%%%
We have the following proposition about a B\"acklund transformation for the generalized soliton solution.
\begin{proposition}  \label{prop 3}
  The generalized solution (\ref{solution I}) with (\ref{def spr}) and the following additional conditions;
\begin{equation}  \label{add cond}
\begin{aligned}
  &0\le q_1\le q_2\le \dots\le q_{N+1}, \\
  &0\le p_1-q_1\le p_2-q_2 \le \dots \le p_{N+1}-q_{N+1},
\end{aligned}
\end{equation}
satisfies a B\"acklund transformation
\begin{equation}  \label{backlund}
  f^n_i +g^{n+\alpha }_{i+\beta } =\max (f^{n+\alpha }_{i+\beta }+g^n_i , f^{n-k+\alpha  }_{i+l+\beta }+ g^{n+k}_{i-l}-A),
\end{equation}
where
$g_i^n$ is the $(N+1)$-soliton solution defined by
\begin{equation}  \label{sol g}
  g_i^n=\max_{\mu_j=0, 1}\Bigl(\sum_{1\le j\le N+1}\mu_js_j(n,i)-\sum_{1\le j<j'\le N+1}\mu_j\mu_{j'}r_j\Bigr),
\end{equation}
and parameters $A$, $\alpha$, $\beta$ satisfy
\begin{equation}  \label{condition alpha beta}
\begin{aligned}
  & A= (k- \alpha )p_{N+1}+(l+\beta )q_{N+1}, \\
  & 0\le \alpha\le k, \qquad 
  -k-l+\alpha \le \beta \le \alpha.
\end{aligned}
\end{equation}
\end{proposition}
Note that the dispersion relations (\ref{disp kdv}) and (\ref{disp toda}) both satisfy the additional condition~(\ref{add cond}).  Therefore the solutions given in the previous section satisfy the B\"acklund transformation (\ref{backlund}) as a special case.\par
  To prove Proposition~\ref{prop 3}, we rewrite $g^n_i$ by
\begin{equation}  \label{newg}
\begin{split}
  g^n_i =\max_{\mu_j=0, 1}\Bigl(&\sum_{1\le j\le N}\mu_j s_j- \sum_{1\le j<j'\le N} \mu_j \mu_{j'} r_j,  \\ 
  &\sum_{1\le j\le N} \mu_j (s_j -r_j)+s_{N+1}- \sum_{1\le j<j'\le N} \mu_j \mu_{j'} r_j, \Bigr).
\end{split}
\end{equation}
Substituting (\ref{solution I}) and (\ref{newg}) into LHS of (\ref{backlund}), we obtain
\begin{equation}
\begin{aligned}
  &f^n_i + g^{n+\alpha }_{i+\beta} = \max_{\mu_j, \nu_j= 0, 1}\Bigl( \sum_{1\le j\le N} (\mu_j +\nu_j) s_j  +\sum_{1\le j\le N}\nu_j(\alpha p_j- \beta q_j) \\
  &\qquad -\sum_{1\le j<j'\le N} (\mu_j \mu_{j'}  +\nu_j \nu_{j'}) r_j  , \\
& \sum_{1\le j\le N} (\mu_j +\nu_j) s_j +  s_{N+1} +\alpha p_{N+1}- \beta q_{N+1} \\
& \qquad + \sum_{1\le j\le N}\nu_j((-k+\alpha ) p_j-(l+\beta ) q_j) - \sum_{1\le j<j'\le N} (\mu_j \mu_{j'}  +\nu_j \nu_{j'}) r_j  \Bigr). \label{eq1}
\end{aligned}
\end{equation}
  Using new parameters $\lambda_i$ and $\sigma_i$ defined by $\lambda_j=\mu_j+\nu_j$, $\sigma_j=-\mu_j+\nu_j$ ($1\le j\le N$), (\ref{eq1}) reduces to
\begin{equation}  \label{eq:1}
\begin{aligned}
  &f^n_i + g^{n+\alpha }_{i+\beta} = \max_{(\lambda_j, \sigma_j)}\Bigl( \sum_{1\le j\le N} \lambda_j s_j  +\frac{1}{2}\sum_{1\le j\le N}(\lambda_j +\sigma_j)(\alpha p_j- \beta q_j) \\
  &\qquad - \frac{1}{2}\sum_{1\le j<j'\le N} (\lambda_j \lambda_{j'} +\sigma_j \sigma_{j'}) r_j, \\
&  \sum_{1\le j\le N} \lambda_j s_j+s_{N+1} +\alpha p_{N+1}- \beta q_{N+1} \\
& +\frac{1}{2}\sum_{1\le j\le N}(\lambda_j+\sigma_j)((-k+\alpha ) p_j-(l+\beta ) q_j)  - \frac{1}{2}\sum_{1\le j<j'\le N} (\lambda_j \lambda_{j'}  +\sigma_j \sigma_{j'}) r_j\Bigr).
\end{aligned}
\end{equation}
Similarly, we have 
\begin{equation}
\begin{aligned}
&f^{n+\alpha}_{i+\beta} + g^n_i = \max_{(\lambda_j, \sigma_j)}\Bigl( \sum_{1\le j\le N} \lambda_j s_j + \frac{1}{2}\sum_{1\le j\le N}(\lambda_j +\sigma_j)(\alpha p_j- \beta q_j)\\
  &- \frac{1}{2}\sum_{1\le j<j'\le N} (\lambda_j \lambda_{j'}  +\sigma_j \sigma_{j'}) r_j, \\
&  \sum_{1\le j\le N} \lambda_j s_j+s_{N+1} +\frac{1}{2}\sum_{1\le j\le N}\lambda_j((-k+\alpha ) p_j-(l+\beta )  q_j) \\
& + \frac{1}{2}\sum_{1\le j\le N}\sigma_j((k+\alpha ) p_j-(-l+\beta ) q_j)  - \frac{1}{2}\sum_{1\le j<j'\le N} (\lambda_j \lambda_{j'}  +\sigma_j \sigma_{j'}) r_j\Bigr), \label{eq:2}
\end{aligned}
\end{equation}
\begin{equation}
\begin{aligned}
&f^{n-k+\alpha}_{i+l+\beta}+ g^{n+k}_{i-l} -A= \max_{(\lambda_j, \sigma_j)}\Bigl( \sum_{1\le j\le N} \lambda_j s_j  +\frac{1}{2}\sum_{1\le j\le N}\lambda_i(\alpha p_j- \beta q_j) -A\\
&+\frac{1}{2}\sum_{1\le j\le N}\sigma_j((-2k+\alpha )p_j - (2l+\beta )q_j) - \frac{1}{2}\sum_{1\le j<j'\le N} (\lambda_j \lambda_{j'}  +\sigma_j \sigma_{j'}) r_j, \\
& \sum_{1\le j\le N} \lambda_j s_j+s_{N+1} +\alpha p_{N+1}-\beta q_{N+1} \\
& +\frac{1}{2}\sum_{1\le j\le N}(\lambda_j+\sigma_j)((-k+\alpha ) p_j-(l+\beta ) q_j)  - \frac{1}{2}\sum_{1\le j<j'\le N} (\lambda_j \lambda_{j'}  +\sigma_j \sigma_{j'}) r_j \Bigr). \label{eq:3}
\end{aligned}
\end{equation}
Note that $\sigma_j$ is redefined by $\sigma_j=\mu_j-\nu_j$ in (\ref{eq:2}) and (\ref{eq:3}).  The former argument of (\ref{eq:1}) is equal to the former of (\ref{eq:2}), and the latter to the latter of (\ref{eq:3}). Hence, (\ref{backlund}) holds if both of the following inequalities hold for any $1\le N'\le N$. 
\begin{equation}
\begin{aligned}
    & \max_{\sigma_j=\pm 1}\Bigl( \sum_{1\le j\le N'}\sigma_j (\alpha p_j- \beta q_j)- \sum_{1\le j<j'\le N'}\sigma_j \sigma_j'r_j  \Bigr)  \\
   & \ge
 \max_{\sigma_j=\pm 1}\Bigl( \sum_{1\le j\le N'}\sigma_j ((-2k+\alpha )p_j-(2l+\beta )q_j)- \sum_{1\le j<j'\le N'}\sigma_j \sigma_{j'} r_j \Bigr) -2A,  \label{max4}
\end{aligned}
\end{equation}
\begin{equation}
\begin{aligned}
  & \max_{\sigma_j=\pm 1}\Bigl( \sum_{1\le j\le N'}\sigma_j ((k- \alpha )p_j+(l+\beta ) q_j)- \sum_{1\le j<j'\le N'}\sigma_j \sigma_j' r_j  \Bigr)  \\ 
 & \ge
 \max_{\sigma_j=\pm 1}\Bigl( \sum_{1\le j\le N'}\sigma_j ((k+\alpha )p_j-(-l+\beta ) q_j)- \sum_{1\le j<j'\le N'}\sigma_j \sigma_{j'} r_j \Bigr) -2\alpha p_{N+1}+2 \beta q_{N+1}.\label{max5}
\end{aligned}
\end{equation}
Since (\ref{max4}) is equivalent to (\ref{max5}) through the transformations $\alpha\to k-\alpha$ and $\beta\to-l-\beta$, we only need to prove (\ref{max5}).  The maximum of LHS of (\ref{max5}) is given by the case of $\sigma_j = (-1)^{N'-j}$, and that of RHS is given by the case of $\sigma_{N'}=1$ and $\sigma_j =(-1)^{N'-j+1}$ $(1\le j\le N'-1)$ respectively\cite{Nagai}.  Thus, the difference between LHS and RHS of (\ref{max5}) is $2(\alpha (p_{N+1}-p_{N'}) -\beta (q_{N+1}-q_{N'}))$ and it satisfies
\begin{equation}
  2(\alpha (p_{N+1}-p_{N'}) -\beta (q_{N+1}-q_{N'})) \ge 2\alpha (p_{N+1}-p_{N'} -(q_{N+1}-q_{N'}))\ge 0.
\end{equation}
Therefore we have proved the proposition.  \par
  Next, we give the B\"acklund transformation in Type II.
\begin{proposition}\label{prop 3-2}
  The generalized solution (\ref{solution II}) with the conditions (\ref{def spr}) and (\ref{add cond}) satisfies the B\"acklund transformation
\begin{equation}  \label{backlund II}
   \tilde f^n_i + \tilde g^{n+\alpha }_{i+\beta } =\max ( \tilde f^{n+\alpha }_{i+\beta }+ \tilde g^n_i -B,  \tilde f^{n-k+\alpha }_{i+l+\beta }+ \tilde g^{n+k}_{i-l}-A),
\end{equation}
where $\tilde g^n_i$ is the $(N+1)$-soliton solution defined by
\begin{equation}  \label{sol g II}
\max \begin{bmatrix}
   |s_1+(-N-1)\displaystyle\frac{r_1}{2}| & |s_1+(-N+1)\displaystyle\frac{r_1}{2}| & \dots & |s_1+(N-1)\displaystyle\frac{r_1}{2}| \\ 
   \dots & \dots & \dots & \dots\\
   |s_{N+1}+(-N-1)\displaystyle\frac{r_{N+1}}{2}| & |s_{N+1}+(-N+1)\displaystyle\frac{r_{N+1}}{2}| & \dots & |s_{N+1}+(N-1)\displaystyle\frac{r_{N+1}}{2}| 
\end{bmatrix},
\end{equation}
Parameters $\alpha $, $\beta $ and $A$ are the same as in (\ref{condition alpha beta}) and $B$ is defined by $\alpha p_{N+1}- \beta q_{N+1}$.
\end{proposition}
Proposition \ref{prop 3-2} is proved after the manner of Proposition \ref{prop 1-2} and Proposition \ref{prop 2-2}.  In particular, using a property of UP,
\begin{equation}
  \max \begin{bmatrix}
  a_{11} & a_{12} & \dots & a_{1N}\\
  a_{21} & a_{22} & \dots & a_{2N}\\
  \dots & \dots & \dots & \dots \\
  a_{N1} & a_{N2} & \dots & a_{NN}
  \end{bmatrix}  +\sum_{1\le j\le N}b_j
 = \max \begin{bmatrix}
  a_{11} +b_1& a_{12} +b_1& \dots & a_{1N}+b_1\\
  a_{21}+b_2 & a_{22} +b_2& \dots & a_{2N}+b_2\\
  \dots & \dots & \dots & \dots \\
  a_{N1} +b_N& a_{N2} +b_N& \dots & a_{NN}+b_N
  \end{bmatrix} ,
\end{equation}
and a formula (\ref{formula}), $\tilde g^n_i$ reduces to
\begin{equation}  
\begin{aligned}
  &\max_{\rho_j=\pm 1}\left(  \max \begin{bmatrix}
   \rho_1 (s_1+(-N-1)\displaystyle\frac{r_1}{2})  &\dots & \rho_1(s_1+(N-1)\displaystyle\frac{r_1}{2}) \\ 
   \dots &\dots & \dots \\
   \rho_{N+1}(s_{N+1}+(-N-1)\displaystyle\frac{r_{N+1}}{2})  &\dots & \rho_{N+1}(s_{N+1}+(N-1)\displaystyle\frac{r_{N+1}}{2}) 
\end{bmatrix}\right) \\
  =&\max_{\rho_j=\pm 1}\left(  \max \begin{bmatrix}
   \rho_1 (s_1-N\displaystyle\frac{r_1}{2})  &\dots & \rho_1(s_1+N\displaystyle\frac{r_1}{2}) \\ 
   \dots &\dots & \dots \\
   \rho_{N+1}(s_{N+1}-N\displaystyle\frac{r_{N+1}}{2})  &\dots & \rho_{N+1}(s_{N+1}+N\displaystyle\frac{r_{N+1}}{2}) 
\end{bmatrix} -\frac{1}{2}\sum_{1\le j\le N+1}\rho_jr_j\right)\\
  =&\max_{\rho_j= \pm 1}\Bigl( \sum_{1\le j\le N+1}\rho_j(s_j-\frac{1}{2}r_j)- \frac{1}{2}\sum_{1\le j<j'\le N+1} \rho_j\rho_{j'} r_j \Bigr)+\frac{1}{2}\sum_{1\le j<j'\le N+1} r_{j'}.
\end{aligned}
\end{equation}
Thus, we can derive through the transformation $\rho_j =2\mu_j-1$,
\begin{equation}
  \tilde f^{n+x}_{i+y} +\tilde g^{n+z }_{i+w } \Bumpeq 2f^{n+x}_{i+y}+2g^{n+z}_{i+w} -\sum_{1\le j\le N}s_j(n+x, i+y)-\sum_{1\le j\le N+1}s_j(n+z, i+w).
\end{equation}
Hence, (\ref{backlund II}) is equivalent to (\ref{backlund}) with a difference of the negligible term $\sum_{1\le j\le N}s_j(n, i)+\sum_{1\le j\le N+1}s_j(n+\alpha, i+\beta)$.\par
%%%%%%%%%%%%%%%%%%%%%%%%%%%%%%%%%%%%%%%%%%%%%%%%%%%%%%%%%%%%%%%%%%%%%
\section{Relation between B\"acklund Transformations for Discrete and Ultradiscrete KdV Equations}
%%%%%%%%%%%%%%%%%%%%%%%%%%%%%%%%%%%%%%%%%%%%%%%%%%%%%%%%%%%%%%%%%%%%%
The $N$-soliton solution of Type II to the discrete KdV equation\cite{Hirotadis1, uBack}, 
\begin{equation}
  F^{n+1}_{i+1}F^{n-1}_i-\delta F^{n-1}_{i+1}F^{n+1}_i-(1-\delta)F^n_{i+1}F^n_i=0,
\end{equation}
is expressed by
\begin{equation}
\begin{aligned}
 F^n_i = \begin{vmatrix}
   \eta_1(n, i) & \eta_1(n+2, i) &\dots & \eta_1(n+2(N-1), i) \\ 
   \dots & \dots &\dots & \dots \\
   \eta_N(n, i) & \eta_N(n+2, i) &\dots & \eta_N(n+2(N-1), i)  
  \end{vmatrix},\label{sol:dKdVf}
\end{aligned}
\end{equation}
where $\eta_j(n, i)$ is defined by
\begin{equation}  \label{4-eta}
  \eta_j(n, i) = c_j \omega_j^n k^i_j+\frac{1}{c_j \omega_j^n k^i_j},  
\end{equation}
with the dispersion relation
\begin{equation} \label{4-dis-dispersion}
  k_j^2 = \frac{1+\omega_j^2 \delta}{\omega_j^2 +\delta}. 
\end{equation}
The B\"acklund transformations for the discrete KdV equation are expressed by
\begin{equation} \label{Back:dKdV}
\begin{aligned}
  &F^n_i G^{n+1}_i = F^{n+1}_i G^n_i /D + F^{n-1}_i G^{n+2}_i/D, \\
  & F^n_i G^{n+1}_{i-1}=\delta F^{n+1}_{i-1} G^n_i /D' + F^{n-1}_{i-1} G^{n+2}_i/D' ,
\end{aligned}
\end{equation}
where $G^n_i$ is the $(N+1)$-soliton solution,
\begin{equation}
\begin{aligned}
 G^n_i =  \begin{vmatrix}
   \eta_1(n-2, i) & \eta_1(n, i) &\dots & \eta_1(n+2(N-1), i) \\ 
   \dots & \dots &\dots & \dots \\
   \eta_{N+1}(n-2, i) & \eta_{N+1}(n, i) &\dots & \eta_{N+1}(n+2(N-1), i) 
  \end{vmatrix},  \label{sol:dKdVg}
\end{aligned}
\end{equation}
and $D$ and $D'$ are defined by
\begin{equation}  \label{4-A}
\begin{aligned}  
  D = \omega_{N+1}+1/\omega_{N+1}, & & D' =k_{N+1}(\omega_{N+1}+\delta /\omega_{N+1}).  
\end{aligned}
\end{equation}
\par
On the other hand, Proposition \ref{prop 3-2} gives the B\"acklund transformations for the uKdV equation, 
\begin{equation} \label{back ukdv1}
\begin{aligned}
  &\tilde{f}^n_i +\tilde{g}^{n+1 }_i  =\max (\tilde{f}^{n +1}_i +\tilde{g}^n_i -p_{N+1}, \tilde{f}^{n -1 }_i+ \tilde{g}^{n+2}_i-p_{N+1}), \\
  &\tilde{f}^n_i +\tilde{g}^{n+1 }_{i-1 } =\max (\tilde{f}^{n +1}_{i-1 }+\tilde{g}^n_i -p_{N+1}-q_{N+1}, \tilde{f}^{n -1 }_{i-1}+ \tilde{g}^{n+2}_i-p_{N+1}+q_{N+1}),
\end{aligned}
\end{equation}
by setting $(k, l, \alpha, \beta )=(2, 0, 1, 0), (2, 0, 1, -1)$ and the dispersion relation
\begin{equation}
  q_j =\min (p_j, 1).
\end{equation}
In particular, we can rewrite $\tilde f^n_i$ and $\tilde g^n_i$ by
\begin{equation}
\begin{aligned}
 \tilde f^n_i = \max \begin{bmatrix}
   |s_1(n, i)| & |s_1(n+2, i)| &\dots & |s_1(n+2(N-1), i)| \\ 
   \dots & \dots &\dots & \dots \\
   |s_N(n, i)| & |s_N(n+2, i)| &\dots & |s_N(n+2(N-1), i)|  
  \end{bmatrix},\label{sol:uKdVf}
\end{aligned}
\end{equation}
\begin{equation}
\begin{aligned}
 \tilde g^n_i =  \max \begin{bmatrix}
   |s(n-2, i)| & |s_1(n, i)| &\dots & |s_1(n+2(N-1), i)| \\ 
   \dots & \dots &\dots & \dots \\
   |s_{N+1}(n-2, i)| & |s_{N+1}(n, i)| &\dots & |s_{N+1}(n+2(N-1), i)| 
  \end{bmatrix}.  \label{sol:uKdVg}
\end{aligned}
\end{equation}
\par
Let us discuss the ultradiscretization of (\ref{Back:dKdV}) and its correspondence to (\ref{back ukdv1}). Introducing the transformations $\delta = e^{-2/\varepsilon }$, $\omega_j = e^{p_j/\varepsilon }$, $k_j = e^{-q_j/\varepsilon }$ and $c_j = e^{c_j/\varepsilon }$, we obtain the ultradiscrete analogues of (\ref{4-eta}), (\ref{4-dis-dispersion}) and (\ref{4-A}) as
\begin{equation}
  \eta_j(n, i) = e^{p_jn-q_ji+c_j}+e^{-p_jn +q_ji-c_j} \quad
\to\quad |p_jn -q_ji +c_j|,
\end{equation}
\begin{equation}  \label{4-dispersion}
 e^{-2q_j/\varepsilon } =\frac{e^{(1-p_j)/\varepsilon } + e^{(-1+p_j)/\varepsilon } }{e^{(-1-p_j)/\varepsilon }  + e^{(1+p_j)/\varepsilon } } \quad
\to\quad q_j =\frac{1}{2}(|p_j+1|-|p_j-1|),
\end{equation}
and
\begin{equation}  \label{4-A2}
  D \to |p_{N+1}|,\qquad D' \to -q_{N+1} +\max (p_{N+1}, -p_{N+1}-2).
\end{equation}
If $p_j$'s are all positive, $q_j$, $D$ and $D'$ in (\ref{4-dispersion}) and (\ref{4-A2}) are reduced to $q_j=\min (p_j, 1)$, $p_{N+1}$ and $p_{N+1}-q_{N+1}$.  However, determinants in (\ref{sol:dKdVf}) and (\ref{sol:dKdVg}) cannot be ultradiscretized directly due to the negative problem. To avoid this missing link, let us assume that ultradiscretization replaces determinant with UP.  By this assumption, the ultradiscrete analogues of $F^n_i$ and $G^n_i$ are replaced with $\tilde f^n_i$ and $\tilde g^n_i$ respectively.  Hence, we have the ultradiscrete analogue of (\ref{Back:dKdV}),
\begin{equation}  \label{back ukdv2}
\begin{aligned}  
  &\tilde{f}^n_i +\tilde{g}^{n+1 }_i  =\max (\tilde{f}^{n +1}_i +\tilde{g}^n_i -p_{N+1}, \tilde{f}^{n -1 }_i+ \tilde{g}^{n+2}_i-p_{N+1}), \\
  &\tilde{f}^n_i +\tilde{g}^{n+1}_{i-1} = \max (\tilde{f}^{n+1}_{i-1} +\tilde{g}^n_i -p_{N+1}+q_{N+1}-2, \tilde{f}^{n-1}_{i-1} +\tilde{g}^{n+2}_i-p_{N+1}+q_{N+1}). 
\end{aligned}
\end{equation}
The latter equation of (\ref{back ukdv2}) seems not to coincide with that of (\ref{back ukdv1}).  However, in the case of $p_{N+1}>1$, $-p_{N+1}-q_{N+1}$ is equivalent to $-p_{N+1}+q_{N+1}-2$ by the dispersion relation.  In the case of $p_{N+1}\le 1$, we can prove 
\begin{equation}  \label{Back:KdVp<1}
  \tilde{f}^{n+1}_{i-1} +\tilde{g}^n_i -p_{N+1}+q_{N+1}-2 < \tilde{f}^{n +1}_{i-1 }+\tilde{g}^n_i -p_{N+1}-q_{N+1}\le \tilde{f}^{n -1 }_{i-1}+ \tilde{g}^{n+2}_i-p_{N+1}+q_{N+1} 
\end{equation}
after the proof in the previous section.  Thus (\ref{back ukdv1}) and (\ref{back ukdv2}) are equivalent each other.
%%%%%%%%%%%%%%%%%%%%%%%%%%%%%%%%%%%%%%%%%%%%%%%%%%%%%%%%%%%%%
\section{Concluding Remarks}
In this article, we consider the generalized soliton solution of two types and give the ultradiscrete soliton equations and the B\"acklund transformation.  The equations are equivalent to the uKdV and the uToda equation in a special case.  The B\"acklund transformation holds under the condition (\ref{add cond}).  The dispersion relations (\ref{disp kdv}) and (\ref{disp toda}) satisfy the condition.  This means that the B\"acklund transformations for the uKdV and the uToda equations are also obtained as a special case.  Furthermore, we discuss the ultradiscretization of the B\"acklund transformations for the discrete KdV equation assuming that determinant is replaced with UP.  Although determinant cannot be ultradiscretized directly in general, their counterpart gives the B\"acklund transformations for the uKdV equation.  
%%%%%%%%%%%%%%%%%%%%%%%%%%%%%%%%%%%%%%%%%%%%%%%%%%%%%%%%%%%%%
%%%%%%%%%%%%%%%%%%%%%%%%%%%%%%%%%%%%%%%%%%%%%%%%%%%%%%%%%%%%%


\begin{thebibliography}{99}
\bibitem{Hirotacont1}
  R.~Hirota:
  Exact Solution of the Modified Korteweg-de Vries Equation for Multiple Collisions of Solitons, 
  J. Phys. Soc. Japan, {\bf 33} (1972) 1456--1458.
\bibitem{Hirotacont2}
  R.~Hirota:
  Exact Solution of the Sine-Gordon Equation for Multiple Collisions of Solitons, 
  J. Phys. Soc. Japan, {\bf 33} (1972) 1459--1463.
\bibitem{Satsuma}
  J.~Satsuma: 
  A Wronskian Representation of $N$-Soliton solutions of Nonlinear Evolution Equations.
  J. Phys. Soc. Japan, {\bf 46} (1979) 359--360.
\bibitem{Freeman}
  N.~C.~Freeman and J.~J.~C.~Nimmo:
  Soliton solutions of the Korteweg-de Vreis and Kadomtsev-Petviashvili equations : the Wronskian technique,
  Phys. Lett. {\bf 95A} (1983) 1.
\bibitem{Hirotadis1}
  R.~Hirota:
  Nonlinear Partial Difference Equations. I. A Difference Analogue of the Korteweg-de Vries Equation, 
  J. Phys. Soc. Japan, {\bf 43} (1977) 1424--1433.
\bibitem{Hirotadis4}
  R.~Hirota:
  Nonlinear Partial Difference Equations. IV. B\"acklund Transformation for the Discrete-Time Toda Equation,
  J. Phys. Soc. Japan, {\bf 45} (1978) 321--332.
\bibitem{Tokihiro}
  T.~Tokihiro, D.~Takahashi, J.~Matsukidaira and J.~Satsuma:
  From Soliton Equations to Integrable Cellular Automata through a Limiting Procedure,
  Phys. Rev. Lett. {\bf 76} (1996) 3247--3250.
\bibitem{Matsukidaira}
  J.~Matsukidaira, J.~Satsuma, D.~Takahashi, T.~Tokihiro and M.~Torii:
  Toda-type cellular automaton and its $N$-soliton solution,
  Phys. Lett. A {\bf 225} (1997) 287--295.
\bibitem{Tsujimoto}
  S.~Tsujimoto and R.~Hirota:
  Ultradiscrete KdV Equation,
  J. Phys. Soc. Japan, {\bf 67} (1998) 1809--1810.
\bibitem{uKdV}
  D.~Takahashi, R.~Hirota:
  Ultradiscrete Soliton Solution of Permanent Type,
  J. Phys. Soc. Japan, {\bf 76} (2007) 104007--104012.
\bibitem{Nagai} 
  H.~Nagai, 
  A New Expression of Soliton Solution to the Ultradiscrete Toda Equation,
  J. Phys. A: Math. Theor. {\bf 41} (2008), 235204(12pp). 
\bibitem{Shinzawa}
  N.~Shinzawa and R.~Hirota:
  The B\"acklund transformation equations for the ultradiscrete KP equation,
   J. Phys. A: Math. Gen. {\bf 36} (2003) 4467--4675.
\bibitem{uBack}
  S.~Isojima, S.~Kubo, M.~Murata and J.~Satsuma: 
  Discrete and ultradiscrete B\"acklund transformation for KdV equation,
  J. Phys. A :Math. Theor. {\bf 41} (2008) 025205(8pp).
\end{thebibliography}
\end{document}